\documentclass{emulateapj}

\usepackage[hidelinks]{hyperref}
\usepackage{aas_macros}

\newcommand{\refchng}[1]{#1}
\newcommand{\mychng}[1]{#1}
\newcommand{\refchngtwo}[1]{#1}

\begin{document}

\title{Ultraviolet \ion{C}{2} and \ion{Si}{3} Transit Spectroscopy and Modeling of the Evaporating Atmosphere of GJ436\MakeLowercase{b}} 

\author{R. O. Parke Loyd \altaffilmark{1}, T. T. Koskinen\altaffilmark{2}, Kevin France\altaffilmark{1}, Christian Schneider\altaffilmark{3}, Seth Redfield\altaffilmark{4}}

\altaffiltext{1}{Laboratory for Atmospheric and Space Physics, 600 UCB, Boulder, CO 80309, USA; \email{robert.loyd@colorado.edu}}
\altaffiltext{2}{Lunar and Planetary Laboratory, University of Arizona, Tucson, AZ 85721, USA}
\altaffiltext{3}{European Space Research and Technology Centre (ESA/ESTEC), Keplerlaan 1, 2201 AZ Noordwijk, The Netherlands}
\altaffiltext{4}{Astronomy Department and Van Vleck Observatory, Wesleyan University, Middletown, CT 06459, USA}

\begin{abstract}
Hydrogen gas evaporating from the atmosphere of the hot-Neptune GJ436b absorbs over 50\% of the stellar Ly$\alpha$\ emission during transit.
Given the planet's atmospheric composition and energy-limited escape rate, this hydrogen outflow is expected to entrain heavier atoms such as C and O.
We searched for C and Si in the escaping atmosphere of GJ436b using far-ultraviolet \textit{HST} COS G130M observations made during the planet's extended \ion{H}{1} transit.
These observations show no transit absorption in the \ion{C}{2} 1334,1335 \AA\ and \ion{Si}{3} 1206 \AA\ lines integrated over [-100, 100] km s$^{-1}$, imposing 95\% (2$\sigma$) upper limits of 14\% (\ion{C}{2}) and 60\% (\ion{Si}{3}) \mychng{depth} on the transit of an opaque disk and 22\% (\ion{C}{2}) and 49\% (\ion{Si}{3}) \mychng{depth} on an extended, highly asymmetric transit similar to that of \ion{H}{1} Ly$\alpha$. 
C$^+$ is likely present in the outflow according to a simulation we carried out using a spherically-symmetric, photochemical-hydrodynamical model.
This simulation predicts a $\sim$2\% transit over the integrated bandpass, consistent with the data.
At line center, we predict the \ion{C}{2} transit depth to be as high as 19\%.
\refchngtwo{Our model predicts a neutral hydrogen escape rate of $1.6\times10^{9}$~g~s$^{-1}$ ($3.1\times10^{9}$~g~s$^{-1}$ for all species) for an upper atmosphere composed of hydrogen and helium.}
\end{abstract}

\section{Introduction}

The hot Neptune GJ436b was recently found to have an escaping atmosphere producing an extended transit that absorbs 56.3\% of its host star's Ly$\alpha$\ (\ion{H}{1}) emission (\citealt{kulow14}; \citealt{ehrenreich15}, hereafter E15; \refchng{\citealt{bourrier16}, hereafter B16}).
This discovery classifies GJ436b as the only Neptune-mass planet among the three planets for which atmospheric escape has been observed in \ion{H}{1}.
The others, HD209458b \citep{vidal03} and HD189733b \citep{lecavelier10} \refchngtwo{(note also a tentative detection for 55 Cnc b; \citealt{ehrenreich12})}, have masses similar to Jupiter.
For these two planets, atoms and ions more massive than H are also entrained in the escaping gas, specifically \ion{O}{1} and \ion{C}{2}, and \ion{Si}{3} in the outflow of HD209458b (\citealt{vidal04,linsky10}\refchng{, though see also \citealt{ballester15} regarding \ion{Si}{3}}) and \ion{O}{1} in the outflow of HD189733b \citep{ben13}.
These species produce transit depths within a factor of two of the \ion{H}{1} transits of those planets.
The escape of heavier species implies that some combination of turbulence (eddy diffusion) and ``drag'' exerted on the heavier species by the escaping hydrogen must be at work (e.g. \citealt{vidal03, vidal04, koskinen10}).

This letter presents a search for transit absorption by \ion{C}{2}, \ion{Si}{3}, \ion{Si}{4}, and \ion{N}{5} in the escaping atmosphere of GJ436b with the G130M grating of the Cosmic Origins Spectrograph (COS) onboard the \textit{Hubble Space Telescope (HST)}. 
We analyzed these data to place upper limits on transit depths, examining simultaneous X-ray data from \textit{Chandra} for information on stellar activity during transit.
Further, we constructed an updated 1D hydrodynamic-photochemical model of GJ436b's thermosphere that predicts H, He, C, and O escape from first principles and compared estimated transit depths with the data upper limits.

\section{Transit Data and Analysis}
\label{sec:data}

\subsection{\textit{HST} Data}
The MUSCLES Treasury Survey \citep{france16p} recently obtained HST observations of GJ436. 
\refchng{Scheduling constraints, including coordinated observations by X-ray observatories, precluded any effort to time the MUSCLES observations with a transit of the star's hot Neptune. 
Nonetheless, three of five HST COS G130M exposures took place during the planet's extended \ion{H}{1} transit. 
The broadband visible-IR (i.e. not extended) transit occurred during the portion of \textit{HST}'s orbit when Earth occulted the target, so it was not captured.}
We extracted lightcurves from the time-tagged photon lists provided for each exposure with an $\sim$500 s cadence (adjusted to evenly divide each exposure) using the procedure outlined in \cite{loyd14}.
In brief, this entailed counting weighted detector ``events'' in the appropriate regions of the spectrum and subtracting an estimate of background counts based on regions offset from the spectral signal trace.
The event weights adjust for fixed pattern noise and detector dead time.

We created lightcurves for every strong line in the COS G130M bandpass that is not contaminated by geocoronal emission: \ion{C}{2} 1334,1335 \AA; \ion{Si}{3} 1206 \AA; \ion{Si}{4} 1393,1402 \AA; and \ion{N}{5} 1238,1242 \AA.
\ion{Si}{4} and \ion{N}{5} are included for completeness, though absorption by these species is not expected due to their high formation temperatures.
In creating lightcurves, we employed two separate velocity ranges over which we integrated emission line flux:
The ``blue integration'', [-100, 30] km s$^{-1}$, includes the velocities predicted by the outflow model of E15, shifted to the heliocentric frame, and the ``full integration'', [-100, 100] km s$^{-1}$, includes all appreciable line flux.

\begin{figure}
\includegraphics{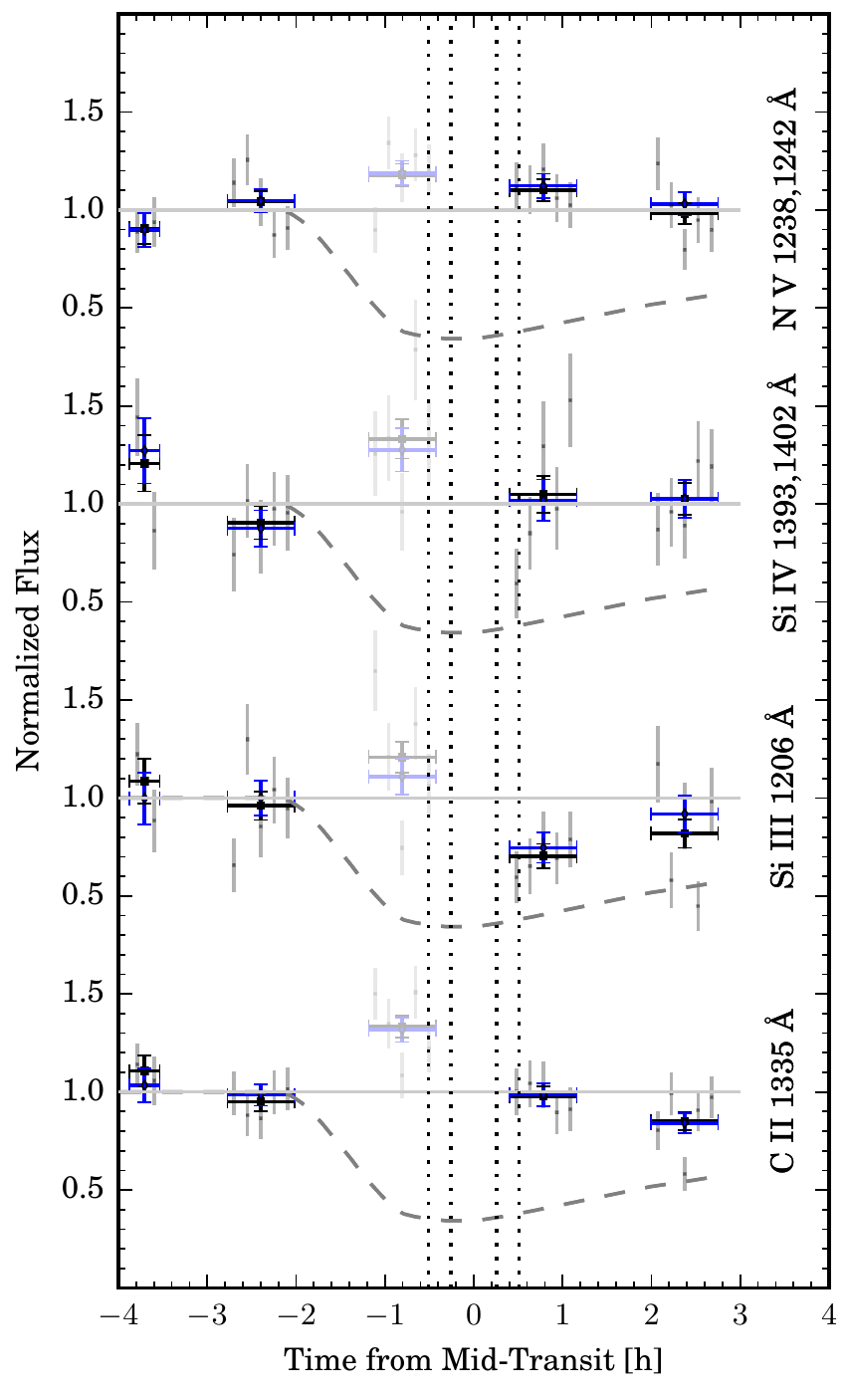}
\caption{Lightcurves of line flux, normalized to the average flux from the first two exposures. 
Blue points represent flux integrated over [-100, 30] km s$^{-1}$, black over [-100, 100] km s$^{-1}$.
Small points integrate flux over $\sim$500 s intervals and large points integrate the full exposures, with horizontal bars showing the exposure duration. 
Dashed gray lines show the E15 \ion{H}{1} transit model.
Dotted vertical lines delimit the contact points of the broadband visible-IR transit.
The third exposure is faint to emphasize that we discarded it from the analysis (Section \ref{sec:activity}).}
\label{fig:lightcurves}
\end{figure}

\begin{figure*}
\includegraphics{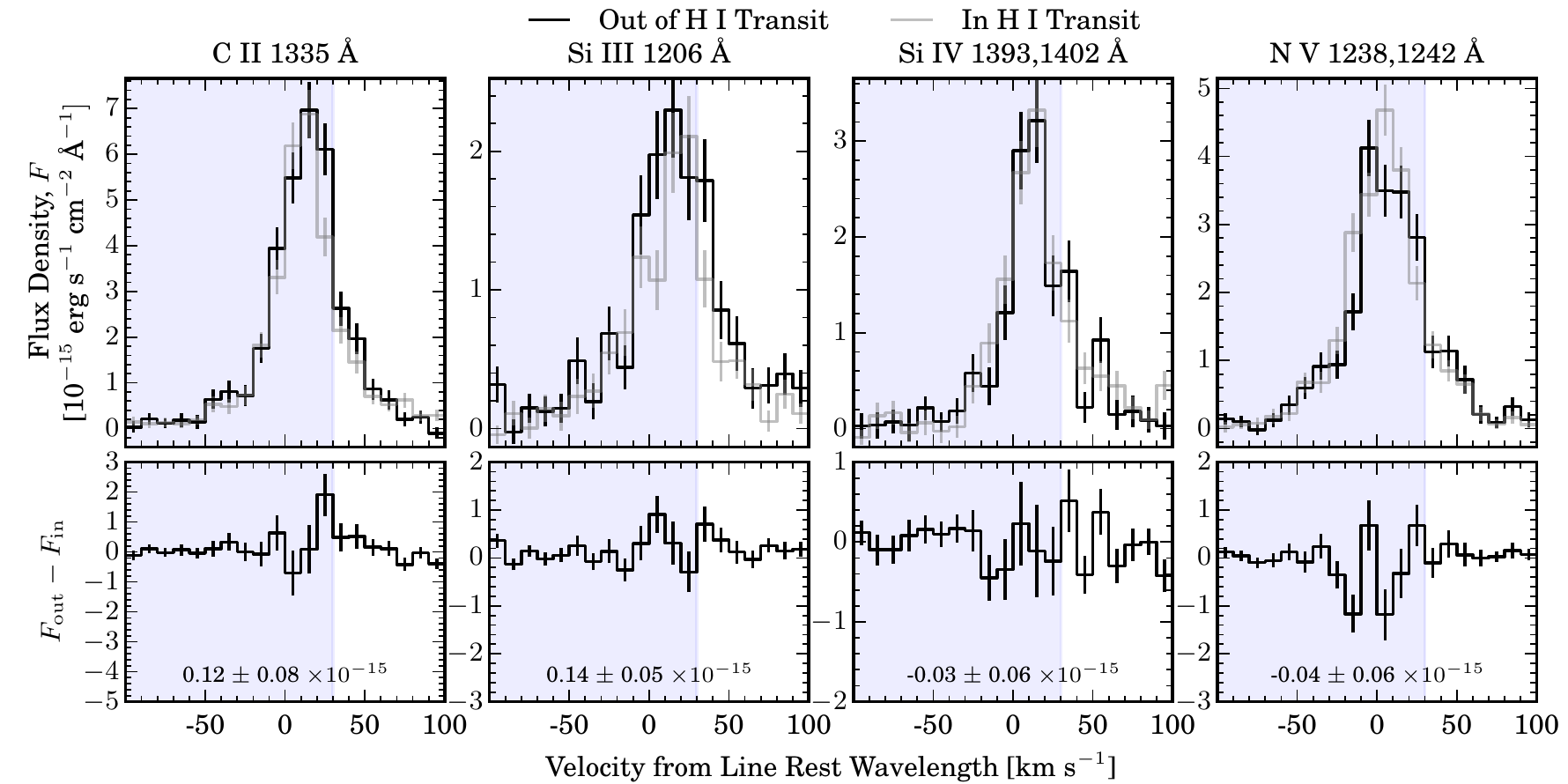}
\caption{Comparison of line profiles prior to (first two exposures) and during the \ion{H}{1} transit (final two exposures).
\refchng{Line wavelengths are listed above each panel, with multiple values indicating coadded lines.
The spectra have been binned to 10~km~s$^{-1}$ per bin for display.}
The plotted wavelength range and the blue shaded area show the full and blue integration ranges used to create lightcurves.
Numbers at the bottom of the lower plot give the flux difference of the full integrations.}
\label{fig:spectra}
\end{figure*}

Figure \ref{fig:lightcurves} displays the lightcurves, with the extended \ion{H}{1} transit fit of E15 overplotted for comparison. 
\refchng{Note that the same group has presented forward modeling with separate fits for each epoch of H I transit data in B16.}
We computed the transit phase of the observations using the ephemeris of \cite{knutson11}.
In addition to integrated time-series, we also searched for velocity-dependent absorption by comparing a spectrum coadded from the first two exposures (pre-transit) to one coadded from the last two exposures (extended \ion{H}{1} transit; Figure \ref{fig:spectra}).

\vspace{2em}
\subsubsection{Correction of a Systematic Wavelength Offset}
\refchng{Due to the instrument configuration, \textit{HST} COS far-ultraviolet spectra have a small gap near their midpoint.
To eliminate this gap, the MUSCLES survey observations employed two central wavelength (CENWAVE) settings, shifting from CENWAVE 1291 to 1318 between orbits 2 and 3. 
This spuriously shifted the measured wavelength of the lines for all 10 of the other MUSCLES targets. 
Left unaddressed, for these observations this shift produces an artificial transit signal at blueward wavelengths for \ion{C}{2} and \ion{Si}{3}.
To correct for this effect, we applied wavelength offsets that produced the minimum $\chi^2$ agreement between line profiles generated before and after the shift. 
These values are  -9 (\ion{C}{2}), -6 (\ion{Si}{3}), 5 (\ion{Si}{4}), and -6 (\ion{N}{5}) km s$^{-1}$.

The CENWAVE change did not significantly influence the measured line flux relative to the statistical noise.
The absolute flux accuracy of the detector is estimated to be 5\% (mainly due to fixed pattern noise; \citealt{coshandbook}). This is reduced to the 1\% level when integrating a line. }

\refchng{\subsection{\textit{Chandra} Data}
Data from the Chandra X-ray Observatory was obtained in tandem with the \textit{HST} data, providing information on stellar activity.
From this data, we created a light curve of the 300~--~2000 eV photon flux (Figure \ref{fig:xcurve}). 
The observation was performed with the ACIS-S detector on board \textit{Chandra}. 
Source photons were extracted within a circle of radius 2 arcsec and we chose a large,  nearby, source-free region to correct for the detector background.}

\subsection{Accounting for Activity-induced Variability}
\label{sec:activity}
Limb brightening \citep{schlawin10}, stellar active regions \citep{llama15}, or any other spatial inhomogeneities in the stellar emission can modify the shape and overall depth of a transit curve.
However, for GJ436b the depth and duration of the \ion{H}{1} transit requires a cloud that covers a large fraction of the stellar disk, diluting effects on the transit from spatial inhomogeneities in stellar emission.

Temporal fluctuations in stellar emission on short timescales contribute to lightcurve scatter. 
These are constrained by two orbits of archival \textit{HST} COS-G130M data (acquired $\sim$14 h before transit and predating the present observations by about 2 years) to \ion{C}{2} $3^{+1}_{-2}$\%, \ion{Si}{3} $<$7\%, and \ion{Si}{4} $<6$\% at the exposure timescale of $\sim$45 min (\citealt{loyd14};
\ion{N}{5} omitted in that work).
We used these values to augment the lightcurve errors when exploring the parameter space of the transit models (Section \ref{sec:signals}).

\refchng{We examined lightcurves for each line at high cadence and discovered a possible flare at $\sim$700~s in the first exposure, visible in \ion{C}{2} and \ion{Si}{3} flux. 
Consequently, we excised the data from 650~--~800~s for all lines.}

\begin{figure}
\includegraphics{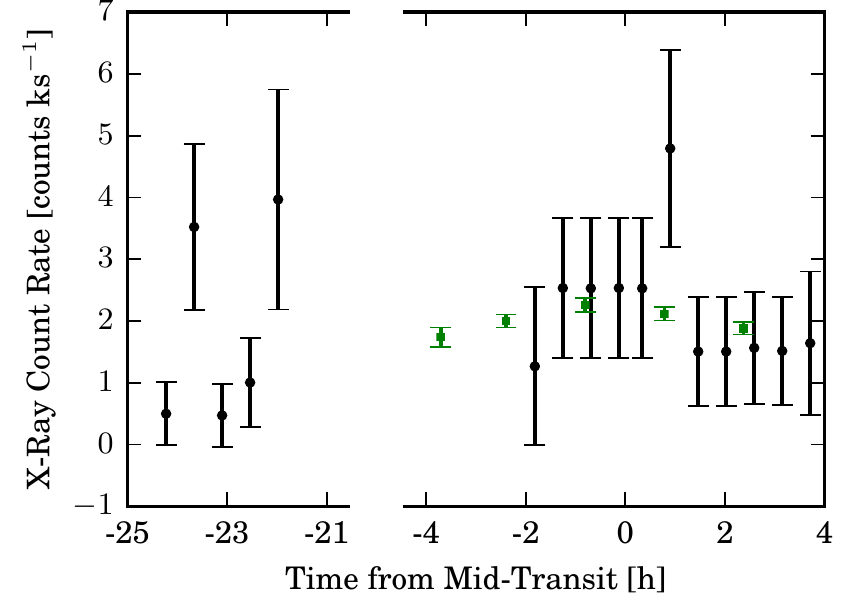}
\caption{Lightcurve of the \textit{Chandra} X-ray observations of GJ436, performed coincident with the \textit{HST} observations. Green squares are the \ion{N}{5} lightcurve normalized to 2 counts ks$^{-1}$, included for reference.}
\label{fig:xcurve}
\end{figure}

Stellar activity is also likely responsible for an enhancement in flux from all four lines throughout the third exposure.
The simultaneity of these enhancements implies a common, physical source.
\refchng{One possible explanation is a flare, but this signal does not exhibit sharp, closely-aligned peaks in \ion{C}{2}, \ion{Si}{3}, and \ion{Si}{4} flux like flares observed on AD Leo and other M stars (\citealt{hawley03, loyd14, france16p, loyd16b} in prep).
Those flares also show responses in \ion{Si}{3} and \ion{Si}{4} many times greater than in \ion{N}{5}, unlike this instance.
No clear signal is seen in the stellar X-ray emission (Figure \ref{fig:xcurve}), though a slight enhancement peaking 1~--~2 h after the FUV lines might be present.}
Regardless of the origin, this signal could mask the beginning of a transit absorption signal, and because it cannot be accurately disentangled from superimposed transit absorption, we omitted it from further analysis.

We speculate that this activity enhancement could result from a star-planet interaction (SPI), such as an accretion stream from the planet. 
The \ion{N}{5} flux, in particular, is well fit by a model wherein a hot-spot traverses the stellar disk at the sub-planetary point\refchng{, but this fit is not significantly better than constant flux.}
In the full MUSCLES sample of stars, \ion{N}{5} emission is strongly correlated with the ratio of planet mass to semi-major axis, suggesting \ion{N}{5} might be a good tracer of SPIs \citep{france16p}. 
\refchng{The hot spot hypothesis could be tested by additional \textit{HST} G130M observations spanning a face-on view of the putative spot (near transit) through its disappearance behind the viewable stellar hemisphere (quadrature).}

\section{Constraints on Transit Signals}
\label{sec:signals}

\label{sec:uplim}
The lightcurves (Figure \ref{fig:lightcurves}) and spectra (Figure \ref{fig:spectra}) show no apparent transit signals.
In order to derive upper limits, we explored the parameter space of two transit models that represent contrasting cases from the zoo of transit shapes that could arise from the many possible ion outflow geometries (e.g. tails, bow shocks, and accretion streams; \citealt{ness65, matsakos15, cauley15, cauley16}). 
As one model we chose the symmetric signal produced by an opaque disk transiting a uniformly illuminated stellar disk, representing scenarios where the ions are decoupled from the \ion{H}{1} outflow and symmetric about the projected planetary disk. 
At another extreme, we chose to scale the highly asymmetric \ion{H}{1} Ly$\alpha$\ transit curve of E15, representing scenarios where ions trail the planet in its orbit, as they would if they were coupled to the \ion{H}{1} outflow.

The parameters of the opaque disk model are the baseline flux level and disk radius.
To these we applied uniform priors for all positive values. 
While the data do not sample the transit of a disk equivalent to GJ436b's broadband visible-IR radius very well (contact points shown as dotted lines in Figure \ref{fig:lightcurves}), the geometry of the transit is such that the disk diameter is a substantial fraction of the path of the transit across the stellar disk.
Consequently, increases in transiting disk radius significantly increase transit duration as well as depth, causing the transit dip to overlap with exposure 4 and allowing that exposure to constrain the model.
We account for the fact that the transit becomes grazing for radii $>$1.8 $R_p$.

The parameters of the H-like model are the baseline flux, ingress time (equivalently transit duration), and depth.
We applied a uniform prior to all positive baseline flux values and all transit depths between zero and unity.  
For ingress time, we use a uniform prior that begins at the end of the first exposure to avoid hyper-extended transits that the data cannot constrain, and ends at the ingress of an equivalent opaque disk to avoid invoking contrived geometries for the absorbing cloud.

We used a Python implementation of the Markov-Chain Monte Carlo (MCMC) method, \textit{emcee} \citep{foreman13}, to sample the parameter space of these models. 
For all model and light curve combinations, no transit is detected.
We report 95\% (2$\sigma$) upper limits for each case in Table \ref{tbl:depths}.
A subset of the H-like models of particular interest consists of transits with duration fixed to that of the \ion{H}{1} transit.
We sampled this subset of models separately and found upper limits on H-like transit depth \mychng{decreased by  $\lesssim$2\%}.

\begin{deluxetable}{lrrrrrr}

\tablewidth{\columnwidth}
\tabletypesize{\scriptsize}
\tablecaption{Upper limit estimates on transit depth. \label{tbl:depths}}

\tablehead{ & \multicolumn{2}{c}{Occulting Disk} & \multicolumn{2}{c}{H-like} \\
\colhead{Line} & \colhead{Blue\tablenotemark{a}} & \colhead{Full\tablenotemark{b}} & \colhead{Blue} & \colhead{Full}}

\startdata
\ion{C}{2} & $<$14\% & $<$14\% & $<$23\% & $<$22\%\\
\ion{Si}{3}\tablenotemark{c} & $<$58\% & $<$60\% & $<$44\% & $<$49\%\\
\ion{Si}{4} & $<$51\% & $<$43\% & $<$31\% & $<$25\%\\
\ion{N}{5} & $<$13\% & $<$12\% & $<$10\% & $<$10\%
\enddata

\tablenotetext{a}{Based on blue integration ([-100, 30] km s$^{-1}$) lightcurve. See Section \ref{sec:data}.}
\tablenotetext{b}{Based on full integration ([-100, 100] km s$^{-1}$) lightcurve.}
\end{deluxetable}

\refchng{The broad integration ranges we used could dilute transit absorption in narrow velocity ranges, so we searched all possible velocity ranges 5~--~100~km~s$^{-1}$ in width for absorption.
To assess the significance of the absorption signals this revealed, we applied the same methodology to 10$^6$ trials of data simulated assuming the null hypothesis of no true absorption, only white noise fluctuations.
None of the absorption signals we found were sufficient to reject the null hypothesis at $>2\sigma$ confidence.}

\section{Model Thermosphere}
\label{sec:model}

\subsection{Model Description}
To interpret the upper limits on transit depth and place constraints on the mass loss rates, we simulated the thermosphere of GJ436b  using an updated version of the 1D escape model of \cite{koskinen13a}. 
This model solves the radial time-dependent equations of continuum, momentum, and energy simultaneously, with the ability to include several ion and neutral species. 
The model treats ionization, recombination, and basic photochemistry of H$_2$, H, He, C, O, H$^+$, H$_2^+$, H$_3^+$, HeH$^+$, C$^+$, O$^+$ and electrons. 
We excluded the related molecules H$_2$O, CO, and CH$_4$ and their chemistry, limiting C and O reactions to ionization, recombination, and charge exchange (excluding reactions with H$_3^{+}$ that create molecular ions). 
The energy equation includes heating by the absorbed stellar XUV radiation, heat conduction, adiabatic cooling and heating, vertical advection, viscous heating and radiative cooling by H$_3^+$ infrared emission and H recombination.  
We included stellar gravity in our calculations to account for Roche lobe overflow effects along the substellar streamline, although these do not significantly affect mass loss from GJ436b.  
\refchngtwo{Magnetic fields are not considered.
Though such fields would influence the flow of ions like C$^+$, no measurements exist for exoplanetary magnetic fields.}

The lower boundary conditions for the model are pressure, temperature, altitude, and composition. 
\refchng{We adopted 1~$\mu$bar as the lower boundary pressure. 
At this pressure level the $T$-$P$ profile derived from secondary eclipse data \citep{line14} gives a temperature of 583~K and an altitude of 3120~km. 
The lower boundary abundances of H$_2$, H, He, C and O are based on solar composition (\citealt{lodders03}\refchng{; see Section \ref{sec:limitations} for discussion}) and a thermal equilibrium ratio of 0.032 for H/H$_2$. }

We placed the upper boundary of the model at 10~$R_p$, well above the L1 point distance (5.8~$R_p$).  
\refchngtwo{The flow is collisional throughout the simulated range of altitudes, as verified using the Knudsen number.}
We impose linearly extrapolated upper boundary conditions on densities, velocity, and temperature that are valid above the Roche lobe and the sonic point.  
We ran the model with the stellar spectrum compiled by the MUSCLES survey \citep{france16p, youngblood16, loyd16} with effective energies ranging from X-rays to the dissociation threshold of H$_2$ near 10 eV. 
This spectrum has an integrated EUV flux (100-911 \AA) of 870 erg s$^{-1}$ cm$^{-2}$ at the 0.0308 AU semi-major axis of GJ436b  \citep{lanotte14}.

\subsection{Model Predictions for Outflow Rate, Composition, and Transit Depths}

Our model predicts an \refchngtwo{escape rate of $1.6\times10^{9}$~g~s$^{-1}$ for neutral hydrogen and $3.1\times10^{9}$~g~s$^{-1}$ for all species.
This exceeds the values of $10^8$~--~$10^9$~g~s$^{-1}$ and 2.5$\pm1\times10^8$~g~s$^{-1}$ estimated by E15 and B16 from model fits to the \ion{H}{1} transit data.  
Their lower escape rate estimates correspond to heating efficiencies below the 11\% efficiency derived from our simulation.
The escape estimate of the B16 model increases to a value agreeing with ours if they impose a fast stellar wind.
However, B16 note that doing so produces a problematic fit to the transit data.}

\begin{figure}
\includegraphics{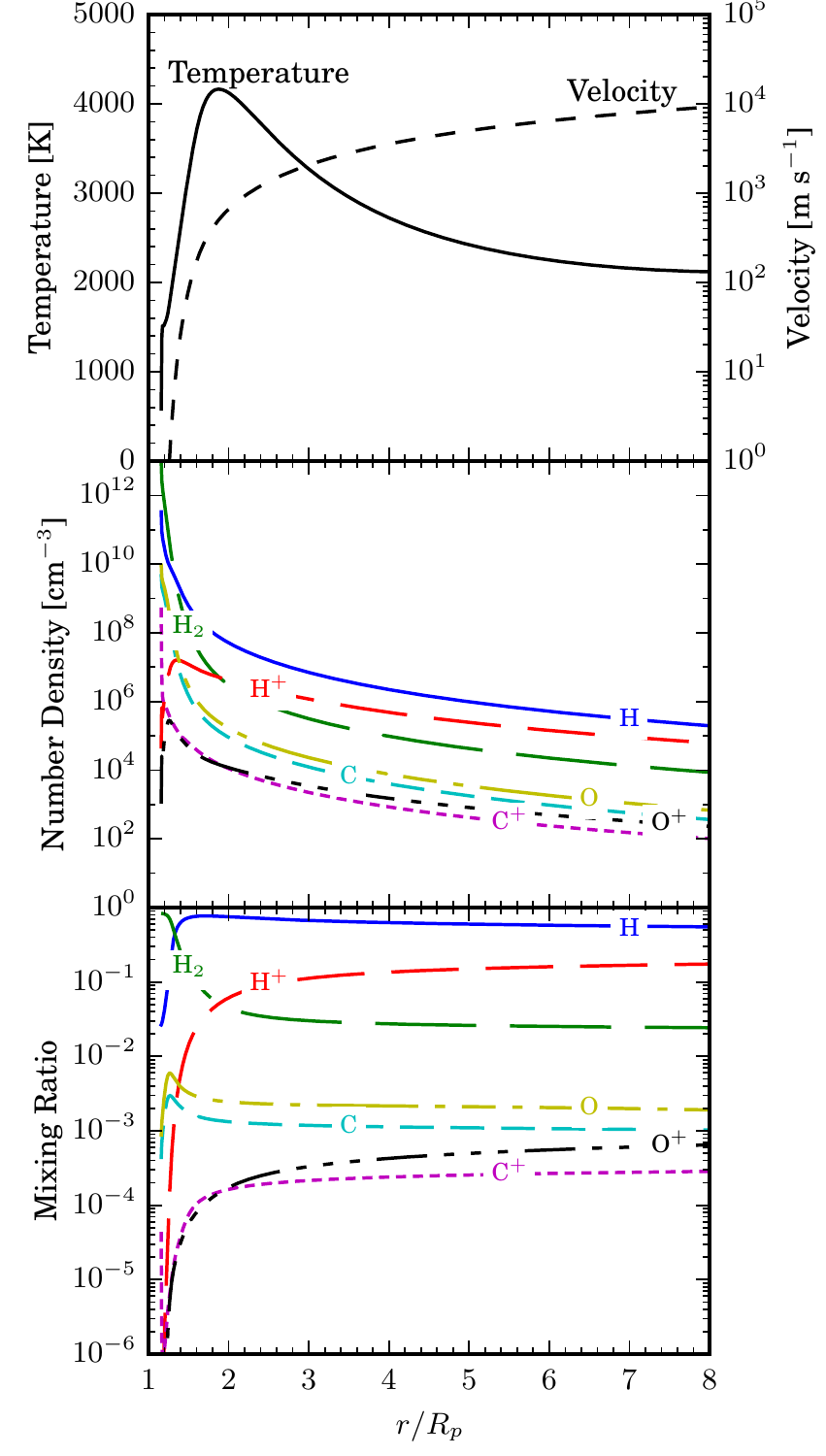}
\caption{Outflow velocity, temperature, and composition of the model thermosphere as a function of height.}
\label{fig:flow}
\end{figure}

\refchng{Figure \ref{fig:flow} gives the temperature, outflow velocity, and composition profiles predicted by our model. 
The outflow velocity reaches 12~km~s$^{-1}$ at the upper boundary, consistent with other simulations of escaping atmospheres \citep{murray09}.
This value is below the 40~--~70~km~s$^{-1}$ required by the B16 model to produce reasonable \ion{H}{1} transit fits.
\mychng{B16 do not give the flow density at their lower boundary (near the Roche lobe), but \cite{bourrier15b} quote a density of $0.4$ cm$^{-3}$ at 4 $R_p$ when describing the precursor to the B16 model, a value that is significantly below our calculations (see Figure \ref{fig:flow}).}}

\refchng{From the density profiles provided by the simulation, rough predictions of transit depth (at mid-transit) for a given line are possible.}
\refchngtwo{For altitudes sampled by the stellar disk at mid-transit but beyond the model boundary, we extrapolate the outflow density profile with an inverse square law.
Alternatively, applying zero density or constant density extrapolations produced roughly a factor of two difference in predicted transit depths.
Beyond the planet's Roche lobe ($\sim$5.8 $R_p$), orbital dynamics, stellar winds, and radiation pressure influence the outflow, producing a broad and asymmetric absorption profile (see E15 and B16).
To approximate modified flows, we explored a parameter space of broadened and shifted \mychng{Voigt} absorption profiles applied to our spherically-symmetric density profile.}

\mychng{Using this procedure to investigate \ion{H}{1} Ly$\alpha$ absorption, we found that absorption profiles blueshited by 20~--~40~km~s$^{-1}$ and given a FWHM of 30~--~50~km~s$^{-1}$ reproduced the $\sim$50\% absorption observed in the [-120, -40]~km~s$^{-1}$ range by E15.}
For \ion{C}{2}, FWHMs of 20~--~80~km~s$^{-1}$ and no velocity shift yielded transit depths in the 1334.5 and 1335.7 \AA\ lines of $\sim$2\%, well below the data-imposed upper limits (Table \ref{tbl:depths}).
At the line centers, the \ion{C}{2} transit depth could be as high as $\sim$19\% if the absorption is narrow (10~km~s$^{-1}$).
\refchngtwo{
Unlike Ly$\alpha$ absorption, absorption at line center for ion lines like \ion{C}{2} is not prevented by ISM absorption.
This permits observations that probe lower levels of planetary outflows than Ly$\alpha$, before the flow is accelerated to velocities outside the ISM absorption profile. 
However, neither the S/N nor the spectral resolution of the present data is sufficient to measure an $\sim$19\% transit over a band as narrow as 10~km~s$^{-1}$.}

An unexpected result of the model is the prediction of significant quantities of neutrals and H$_2$ in the outflow.
H$_2$ at the predicted temperatures would have substantial populations in the $v = 1$ and 2 vibrational levels that have a dense forest of electronic transition lines in the FUV.
This absorption could be measured against strong stellar emission lines, like Ly$\alpha$, \ion{O}{6} 1032 \AA, and \ion{C}{3} 977 and 1175 \AA\ by future missions with enhanced resolution and sensitivity in the FUV.

\subsection{Limitations of the Model}
\label{sec:limitations}

The model's omission of H$_2$O, CO, and CH$_4$ could be important for GJ436b, as radiative cooling by these molecules could significantly reduce the mass loss heating efficiency of the atmosphere and thus lower the mass loss rate. 
This simplification was justified when applying the same model to hot Jupiters, such as HD209458b \citep{koskinen13a}, because detailed photochemical models predicted these molecules dissociate before reaching outflow altitudes \citep{moses11, lavvas14}, but this might not be the case for GJ436b \citep{moses13}.
If substantial populations of these molecules are present, \refchng{they could be transported into the thermosphere through turbulent mixing or collisional ``drag'' by rapidly escaping hydrogen, as we show in the following paragraphs.

Turbulent mixing can be parameterized in the form of an eddy diffusion coefficient, $K_{zz}$.
\mychng{We find that if $K_{zz} \gtrsim10^{8}$~cm$^2$~s$^{-1}$,} H$_2$O can be transported above the 1~$\mu$bar level.
Estimates of $K_{zz}$ for hot Jupiters (e.g. \citealt{moses13}) exceed this value.
If similar values apply to GJ436b, then turbulence will transport molecules into the upper atmosphere. 

Collisional drag can effectively transport any species with a mass below the crossover mass,
\begin{equation}
m_{xo} = m_{H} + \frac{k T \dot{M}}{4 \pi nD G M_p \mu},
\end{equation}
where the quantities involved are the mass of hydrogen, $m_H$; Boltzmann constant, $k$; gas temperature, $T$; mass loss rate, $\dot{M}$; coefficient of diffusion between the particle and hydrogen, $nD$; gravitational constant, $G$; planet mass $M_p$; and mean molecular weight, $\mu$ \citep{hunten87}.
Using the properties of the simulated outflow and the approximation
\begin{equation}
nD = 1.52\times10^{18}\sqrt{\frac{T}{\mathrm{K}} \left( \frac{1}{m_{xo}/\mathrm{amu}} + 1 \right)}\ \mathrm{cm^{-1} s^{-1}} 
\end{equation}
\citep{koskinen13b}, we find a crossover mass of 28 amu, well above the mass of H$_2$O and CH$_4$.
Therefore, these molecules could be dragged along with the outflow.}

\mychng{The presence of molecules could be enhanced by super-solar abundances of C and O. 
A super-solar metallicity is necessary to explain the ratio of 3.8 \micron\ to 4.5 \micron\ planetary emission measured via secondary eclipse \citep{stevenson10, madhusudhan11, moses13, lanotte14}.
However, for this work we assumed solar metallicity.
The enhanced molecular cooling at higher metallicities could produce a shallower \ion{C}{2} transit if it reduces the escape rate by an amount that overcomes the increased mixing ratio of C$^+$, producing lower C$^+$ column densities at a given altitude.}

The addition of H$_2$O, CO, and CH$_4$ to the model and the exploration of higher metallicities is beyond the scope of this Letter, but provides motivation for future work.

\section{Discussion and Conclusions}
\label{sec:discussion}
We found no evidence of transit absorption by \ion{C}{2} or \ion{Si}{3} in \textit{HST} COS G130M observations of GJ436b.
The non-detection of a \ion{Si}{3} transit is likely due to blockage of upward Si transport by condensation traps, namely clouds of forsterite (Mg$_2$SiO$_4$) and enstatite (MgSiO$_3$).
These species have condensation curves \citep{fortney05} that cross the $T$-$P$ profile for GJ436b retrieved by \cite{line14}.
Further, we estimate that Si bearing molecules are too heavy to be dragged into the upper atmosphere by escaping H.
By these same arguments, Mg, another species observed in extended hot-Jupiter atmospheres (HD209458b and WASP-12b; \citealt{fossati10, vidal13, bourrier14, bourrier15a}), is unlikely to be present in the upper atmosphere of GJ436b.

The data impose upper limits on \ion{C}{2} transit depth that are well below the \ion{H}{1} Ly$\alpha$\ transit, unlike the hot Jupiter HD209458b \citep{vidal04, linsky10}.
To test this result, we employed a simulation of GJ436b's thermosphere to predict \ion{C}{2} populations and found them significant, yet too low to produce sufficient transit absorption for detection with the present dataset:
The data-imposed upper-limits of 14\% (occulting disk transit) and 22\% (H-like transit) substantially exceed the model-predicted transit depth of $\sim$2\%.
The simulation further predicted densities of \ion{H}{1} capable of producing the $\gtrsim$50\% transit observed in previous data (E15) and an escape rate of $3.1\times10^{9}$~g~s$^{-1}$ (all species) that implies a mass loss efficiency of 11\%. 

\refchngtwo{The escape rate we estimated exceeds the 2.5$\pm1\times10^8$~g~s$^{-1}$ estimate of B16.
Further, the predicted densities and outflow velocities of these models differ significantly. Future work is needed to resolve the discrepancy between the Monte-Carlo forward modeling (``top-down") approach of B16 and our photochemical-hydrodynamical (``bottom-up") approach. A thorough understanding of escape from GJ436b is of particular importance given that it is currently the sole Neptune-mass planet with a known escaping atmosphere.}

\section{Acknowledgments}
The authors made use of Astropy code contributed by Stuart Little to compute barycentric light travel time corrections and are thankful for the large amount of effort this saved. The data presented here were obtained as part of the HST Guest Observing program \#13650. This work was supported by NASA grants HST-GO-13650.01 to the University of Colorado at Boulder. 


\end{document}